\documentclass[aps,pra,twocolumn]{revtex4}
\newcommand {\be}{\begin{equation}}
\newcommand {\ee}{\end{equation}}
\newcommand {\bey}{\begin{eqnarray}}
\newcommand {\eey}{\end{eqnarray}}

\usepackage{mathbbol}
\usepackage{epsfig}
\usepackage{amssymb}
\usepackage{amsfonts}
\usepackage{amsmath}
\usepackage{enumerate}
\usepackage{amsthm}
\usepackage{pgfplots}
\usepackage{tikz}
\usepackage{pgf}

\usepackage{hyperref}

\newtheorem{definition}{Definition}
\newtheorem{theorem}{Theorem}
\theoremstyle{definition}

\newtheorem{lemma}{Lemma}
\newtheorem{model}{Model}

\newtheorem{assumption}{Assumption}

\newcommand{\mathsym}[1]{{}}
\newcommand{\unicode}[1]{{}}

\begin{document}

\title{Quantum-foundational implications of information erasure upon measurement}
\author{Alberto Montina, Stefan Wolf}
\affiliation{Facolt\`a di Informatica, 
Universit\`a della Svizzera italiana, 6900 Lugano, Switzerland}
\date{\today}

\begin{abstract}
A projective measurement cannot decrease the von Neumann entropy if the outcome is ignored.
However, under certain sound assumptions and using the quantum violation of Leggett-Garg 
inequalities, we have previously demonstrated that this property is not inherited by a 
{\it classical} simulation of such a measurement process.
In the simulation, a measurement erases prior information by partially resetting the system, 
suggesting that the quantum-state update following a measurement cannot be entirely epistemic.
The erasure of information has been proved by assuming that the maximally mixed quantum state 
corresponds to maximal ignorance of the classical state. A more intricate proof employed 
the weaker hypothesis that the entropy is finite at some stage of the simulation. In this
paper, we focus on the quantum-foundational implications of this theorem. We first provide
a simple proof by directly using the second hypothesis. Second, we identify information 
erasure as the mechanism breaking the time symmetry in ontological theories. This symmetry
break has been previously proved by Pusey and Leifer. 
Third, we show that information erasure and, thus, 
symmetry break can be avoided by employing a branching {\it \`a la} many-worlds theory.
The information flow and the time asymmetry are transferred to the measurement devices and 
the subsequent comparison of results, which inherently involve time-asymmetric processes. 
Thus, causality and the absence of information erasure suggest that measurements have multiple 
actual outcomes. Similarly, Deutsch and Hayden argued that Bell’s theorem leads to the 
same conclusion if locality is given for granted. We conclude by showing that the problem of 
the clumsiness loophole in an experimental Leggett-Garg test of macrorealism is mitigated 
by the information-erasure theorem.
\end{abstract}
\maketitle

\section{Introduction}

In the Copenhagen interpretation of quantum theory, there is a strange separation
between the fuzzy quantum realm and the sharp macroscopic reality, known as the
Heisenberg cut. While this separation does not have practical impacts because
of decoherence in open systems, it raises a conceptual question: Does quantum coherence 
really hold in closed systems
at the macroscopic level, as in the de Broglie-Bohm~\cite{bohm,bohm2} and 
many-worlds~\cite{everett} theories, or does the macroscopic reality emerge from 
a breakdown of unitary evolution at some level? In the context of this question, 
Leggett and Garg (LG) proposed a criterion for 
experimentally testing the emergence of macroscopic realism from a breakdown 
of unitarity~\cite{LG}.  They first formulated two assumptions which are
justified by models of wave-function collapse. In these models, the collapse of 
the  wave-function occurs
with some probability rate as if some observable $\hat A$ were measured. 
Thus, if $\hat A$ is actually measured immediately after a spontaneous collapse,
then the outcomes are consistent with two assumptions:
First, we can assume that the observable $\hat A$ had a definite value $a$ 
which a measurement reveals. Second, the measurement does not modify the 
statistics of subsequent measurements. The two assumptions are called by Leggett 
and Garg (A1) {\it macroscopic realism} and (A2) {\it noninvasive measurability}.
Taking them as the minimal requirement for the emergence of a macroscopic reality,
Leggett and Garg derive inequalities that are violated by quantum systems undergoing a 
unitary evolution between measurements. The experimental violation of the inequalities
would be a proof of the failure of one of the two assumptions.

As illustrated by the de Broglie-Bohm theory and highlighted in Ref.~\cite{maroney}, 
the breakdown of unitarity is not necessary for a realistic theory, also known
as an {\it ontological theory}~\cite{harrigan}.
Roughly speaking, ontological theories can be seen as classical simulations of quantum 
processes, except that they are intended as physical theories providing a ‘realistic’ 
picture of the processes underlying macroscopic observations. In
this paper, we consider ontological models of a quantum system undergoing
sequential measurements under the assumption of {\it unitarity} and 
{\it causality}, (being meant as `no influence 
from the future to the past'). Thus, we interpret the violation of the LG 
inequalities as a fealure of Assumption~(A2), rather than a fealure of realism. 

It is not a surprise
that quantum measurements are invasive. Indeed, a measurement
modifies the probabilities of the outcomes of a subsequent noncommuting
measurement. However, in Ref.~\cite{montina_isit}, we have shown that
the violation of LG inequalities implies more than a mere break of assumption~(A2).
While a projective quantum measurement does not decrease the entropy if the outcome
is ignored, the perturbation induced by the measurement cannot be reproduced by
a classical simulation without a partial reset of the classical state of the measured system.
Thus, using the conventional terminology of Ref.~\cite{croucher}, the measurement 
{\it erases information}. {\it Information erasure} is implied by the quantum violation 
of the LG inequalities under the hypothesis that the maximally mixed quantum state corresponds 
to maximal ignorance of the underlying classical state. We have also provided a more intricate 
proof by employing the weaker hypothesis that the entropy of the system is finite at some 
stage of the simulation~\cite{montina_isit}. In Ref.~\cite{montina}, it was shown that the 
erasure of just one bit suffices to account for the outcome statistics of a two-state
system, the measurements being performed at two arbitrary times.

In Ref.~\cite{montina_isit}, we primarily focused on the information-theoretic problem 
of classically simulating sequential quantum measurements. In this paper, we explore 
the quantum-foundational implications of information erasure within ontological theories. 
We present four main results. First, we provide a simple proof of information erasure 
by directly using the second, weaker hypothesis from Ref.~\cite{montina_isit}. Second, 
we identify {\it information erasure} as the mechanism responsible for breaking 
time symmetry in ontological theories. This breaking has been previously proven
by Leifer and Pusey~\cite{leifer} under the assumption of causality.
Third, we show that both our 
theorem and the Leifer-Pusey theorem can be circumvented within a framework
{\`a la} many-worlds theory.
Specifically, we introduce a model with two parallel coexisting
realities (called {\it instances} in Ref.~\cite{montina_branch}) 
that offers a fully time-symmetric 
description of a scenario in which a qubit, initially in a 
maximally mixed state, is measured at two different times. The time asymmetry 
established in Ref.~\cite{leifer} is instead transferred to the measurement 
devices and 
the subsequent comparison of results, which inherently involve time-asymmetric 
processes. This model is the temporal analogue of the {\it local} two-instances 
model in Ref.~\cite{montina_branch}, which simulates quantum correlations 
between two maximally entangled qubits. 
Finally, we show that the information-erasure theorem helps mitigate 
the clumsiness loophole in an experimental Leggett-Garg test of macrorealism.

Despite their issues~\cite{kochen,bell,pbr,montina_isit,leifer}, 
ontological theories can provide insights into potential flaws in their underlying 
assumptions. Two possible assumptions under scrutiny are causality and the existence of
a single macroscopic reality. The former has been questioned in Ref.~\cite{leifer}.
By discarding the latter, our model shows that both information erasure and time symmetry 
breaking can be avoided. Similarly, it has been previously argued 
that Bell's theorem does not conflict with locality if multiple actual realities
are assumed~\cite{deutsch2}. Even contextuality~\cite{kochen} and the 
Pusey-Barrett-Rudolph theorem~\cite{pbr}, which rules out $\psi$-epistemic ontological
theories, may be circumvented through a framework akin to the many-worlds 
theory~\cite{montina_branch}. This claim is supported by a connection between 
information erasure and the debate on $\psi$-epistemic theories~\cite{harrigan,pbr,barrett}. 
The reset of the ontic state would imply that the quantum-state update following a 
measurement cannot be purely epistemic, conflicting with one motivation for 
$\psi$-epistemic theories. However, this conflict is
removed in our two-instances model.
Moreover, information erasure is linked to preparation contextuality,
another quantum-foundational concept defined in Ref.~\cite{spekkens}, 
suggesting that contextuality may be circumvented in the same manner.

The paper is organized as follows. 
In Sec.~\ref{clas_model_sec}, we introduce a general ontological model 
describing projective measurements and unitary evolutions.
In Sec.~\ref{leg_garg_sec}, we show that the violation of Leggett-Garg inequalities
implies a flow of information from the past to the future, even if
signaling is not allowed. With this premise,
in Sec.~\ref{inf_erasure}, we prove the theorem on {\it information erasure}
under the hypothesis that there is a quantum state compatible with a 
distribution of finite entropy at the ontological level. Information erasure is
physically interpreted in Sec.~\ref{sec_interpr} as entropy flow from the system to 
the low-entropy measuring device. We conclude the section by discussing in 
details the relation between information erasure and Spekkens' preparation 
contextuality~\cite{spekkens}.
A link between information erasure and contextuality has been also discussed in
Ref.~\cite{catani}.
Illustrations of the theorem are presented in Appendix~\ref{sec_dBB}, where the 
Beltrametti-Bugajski model~\cite{beltrametti} and the de Broglie-Bohm theory 
are discussed.
In Sec.~\ref{sec_time_symm}, we discuss the relation between information erasure and 
the theorem of Leifer and Pusey. Time symmetry is the 
temporal version of Bell’s locality assumption and comes from  a kind of ‘no fine-tuning 
principle’. Leifer and Pusey prove that time simmetry is in conflict with 
causality. We identify information erasure as the mechanism leading to the break of 
the time symmetry and, thus, of the ‘no fine-tuning principle’. This mechanism
suggests the conceptual step taken in Sec.~\ref{many-worlds_sec}, where we introduce 
the two-instances model that evades the information-erasure theorem and, thus, the 
theorem by Leifer and Pusey. Finally, in Sec.~\ref{sec_clumsy}, we discuss
the `clumsiness loophole' in Leggett-Garg tests and argue that the information-erasure
theorem mitigates this loophole.

\section{Ontological theories}
\label{clas_model_sec}
An ontology for quantum theory is formulated whenever
it is specified which aspects within the theory should be considered as elements of reality.
Thus, an ontological assessment is mandatory for any physical
theory. The Copenhagen interpretation holds that macroscopic events that we can
experience constitute elements of reality. This minimal requirement is related to the assumption 
of {\it absoluteness of observed events}, introduced in the context of Wigner's friend thought
experiments~\cite{wiseman}.
However, the interpretation avoids establishing 
a causal connection between events through a continuous sequence of underlying
states of reality, commonly referred to as {\it ontic states}. In this view, the quantum state 
is merely a mathematical construct rather than a fundamental element of reality. 
The Copenhagen interpretation can be considered as a kind of minimal ontological theory,
in which factual events are immersed in a `fog' of indeterminacy.

In the strict sense,
{\it ontological theories}~\cite{spekkens0} aim to provide the causal connection missing
in the Copenhagen interpretation. The paid price is the introduction of elements of reality 
which are not directly observed.  At each time, a quantum system is represented 
through an ontic state, say $\lambda$, which is an element within an 
ontological space $\Lambda$. Generally, it is assumed that
this state evolves deterministically in a closed system, which mirrors 
the reversibility of unitary evolutions. In particular, we will give for granted that
the evolution of a closed system is deterministic and volume preserving. 
The most direct route toward such a theory is to interpret the wave function as
a physical entity. Ontological theories
of this kind, in which $\lambda$ contains the wave-function, are known as $\psi$-ontic.
Probably, this interpretation of the quantum state is shared, at least implictly, by many 
physicists and is at the basis of the many-worlds theory. Setting aside that 
theory, the sole ontology of the wave-function ($\lambda=|\psi\rangle$) does not 
suffice to account
for our well-defined experience because of the linearity of the Schr\"odinger 
equation. A simple resolution to this issue is to break the reversibility of
the evolution at the macroscopic level, such as in collapse theories, 
which Leggett and Garg aimed to test through their inequalities. 
Another approach, which avoids the break of unitarity, 
introduces additional variables specifying, in particular, the
actual values of macroscopical observables. This realistic completion of quantum
theory is consistently obtained in de Broglie-Bohm theory, in which the wave-function
is supplemented by the positions of the particles. 

Since the de Broglie-Bohm theory treats the wave function as a real, physical entity, 
it has been argued that it exhibits a branching structure similar to that of the 
many-worlds theory (see Ref.~\cite{brown} and references therein). The key distinction 
is that, in de Broglie-Bohm theory, the positions of particles label the branch 
that is actually experienced, while other branches, although real, remain "empty" 
of observers. An alternative route to a realistic picture
of quantum phenomena which avoids attributing reality to all the branches of 
the universal wave-function is offered by so-called $\psi$-epistemic ontological 
theories~\cite{spekkens0}.
Like collapse theories and de Broglie-Bohm theory, there is an ontic state $\lambda$
which describes the actual state of affairs of the system at each time 
and this state contains, in particular, the information about macroscopic observables. 
However, the full information about the quantum state is encoded in the probability 
distribution of the ontic state rather than in each instance of the ontic state. In 
this respect, $\psi$-epistemic theories share with the Copenhagen interpretation the 
view that quantum states are just mathematical constructs, but they differ in what they
consider real entities.

Since ontological theories aim to provide a unified picture of nature without a
distinction between system and observer, they refer to closed systems.
In the following, we introduce an ontological model for a system that is sequentially measured 
by an external device. The first assumption 
we make is that the ontic state $\lambda$ of the system belongs to a fixed ontological space 
$\Lambda$, which is independent of the measurements performed on the 
system. This mirrors the operational framework of quantum theory, where the state of a 
system is represented by a reduced density operator, or equivalently, a mixture of pure 
quantum states in a fixed Hilbert space. We also assume that the ontological space is 
measurable, with a finite measure. When a system is measured, the effect of the 
measurement on the ontic state is modeled as a stochastic process. The model is as follows.
\begin{model}{Ontological model of a (finite-dimensional) quantum system undergoing projective 
measurements}
\newline
\begin{enumerate}
\label{causal_model}
\item At each time, the ontic state of the system is some element $\lambda$ of 
a measurable {\it ontological space} $\Lambda$ with a finite measure (volume).
Furthermore, there is a surjective map
\be\label{sur_map}
\rho(\lambda)\in\Omega \rightarrow \hat\rho\in\mathcal{D},
\ee
where $\Omega$ is a convex set of probability distributions on $\Lambda$ and
$\mathcal{D}$ is the space of density operators. At each time, $\rho$ is 
mapped to the quantum state.
\item A unitary evolution corresponds to a volume-preserving 
transformation in $\Lambda$.
\item The execution of the measurement $\hat A$ modifies an incoming value 
$\lambda^{in}\in \Lambda$ to a outgoing value
$\lambda^{out}$ according to a conditional probability $\rho_{\hat A}(\lambda^{out}|\lambda^{in})$.
The measurement outcome $a$ is generated with a conditional probability 
$\rho(a|\lambda^{in},\lambda^{out})$
[This requirement is less restrictive than both Assumption~(A1) and that in 
Ref.~\cite{montina_isit}, where the outcome $a$ is conditioned only on
$\lambda^{in}$].
\item The value of $\lambda$ is statistically independent of the execution of
future measurements and unitary evolutions (causality).
\end{enumerate}
\end{model}
Model~\ref{causal_model} defines building blocks for describing 
any sequence of measurements on a unitarily evolving quantum system. 

Let us discuss and characterize each property defining Model~\ref{causal_model}.\newline
{\bf Property 1}. The state of the system at each time is encoded by some 
element $\lambda$ in a space $\Lambda$. Employing a popular term in quantum foundation, 
we have called $\lambda$ an {\it ontic} state. A pure quantum state does not necessarily 
determine uniquely the ontic state (unlike in collapse theories, where it does), but 
corresponds to some probability distribution $\rho(\lambda)$. 
Due to preparation contextuality~\cite{spekkens}, each quantum state may correspond 
to a multitude of probability distributions, so that the map $\hat\rho\rightarrow\rho(\lambda)$
is not single-valued. Rather, we define a set of probability distributions $\Omega$ on which 
the surjective map~(\ref{sur_map}) is defined.

The ontological space must have infinite elements, as stated by the {\it excess
baggage theorem} \cite{hardy}. Moreover, it is uncountably infinite. which is 
implied by the short-memory (Markovian) evolution employed in the classical 
model~\cite{montina_dim,montina_dim2}. The ontological space may have disjoint parts. 
For example, a point may be determined by a set of continuous variables and some additional
finite number of bits. Since the space is uncountably infinite, we have to define
a measure on it. We assume that the volume of $\Lambda$ is finite. For example, 
the space may be a hypersphere or a compact subset of an Euclidean space. This is
the case in objective-collapse theories and the Beltrametti-Bugajski model~\cite{beltrametti},
in which the ontological space is the space of normalized quantum states.
Since the ontological space has {\it finite} volume, the differential entropy 
\be
H(\rho)\equiv  -\int d\lambda\rho(\lambda)\log\rho(\lambda)
\ee
is upper-bounded, however it might be equal to $-\infty$. Later, we will 
assume that 
the entropy of the ontic state is finite for some quantum state. This may be justified by
(completely) $\psi$-epistemic models~\cite{montina_comm_epist0,montina_comm_epist}.
\newline
{\bf Property 2}.
Since unitary evolutions form a group, it is reasonable to assume that they
are associated with transformations of the ontic state which preserve the volume. Thus, 
the entropy is constant under unitary transformations of the quantum state. 
\newline
{\bf Property 3}.
When a measurement is performed, an outcome is 
generated with a probability depending on the ontic state  prior to and after 
the measurement, say $\lambda^{in}$ and $\lambda^{out}$. 
In Ref.~\cite{montina_isit}, we assumed that the outcome depends only on 
$\lambda^{in}$. A stronger assumption would 
demand that the outcome is determined by $\lambda^{in}$  [equivalent
to Assumption~(A1)].  These differences do not affect our conclusions.
The measurement modifies the ontic state according to some transition probability. 
The measurement acts on the system according to the conditional probability
$$
\rho_{\hat A}(\lambda^{out}|\lambda^{in}).
$$
Here, we give for granted that the function $\rho_{\hat A}$ depends only on the measurement
procedure, regardless of the time at which it is executed. That is, we assume that 
all the memory on the previous history is encoded in the present state of the system.
We consider a unique procedure for the measurements, so that  the conditional probability
$\rho_{\hat A}(\lambda^{out}|\lambda^{in})$ is fixed.
Note that this characterization of a measurement generally requires a break of the
time symmetry, which 
mirrors the symmetry break of the corresponding quantum process. A projective 
measurement generally increases the von Neumann entropy (after tracing out the
outcome). 
\newline
{\bf Property 4.}
We assume that the ontic state is uncorrelated to any 
future choice. This requirement is necessary to prove the information erasure. We will not
consider possible weaker hypotheses admitting retrocausality, which may also require a different
description of measurements in the model.

\section{Leggett-Garg inequalities and information flow}
\label{leg_garg_sec}

There is a formal analogy between LG inequalities and Bell's inequalities, which leads
to the conclusion that there is a flow of information from the past to the future
if the former is violated. Crucially, this occurs even if signaling is not allowed. 
We will show that in the framework of the ontological Model~\ref{causal_model}.

The LG inequalities~\cite{emary} refer to a scenario in which a 
measurement $\hat A$ is executed at two times $t_k$ and $t_l$ chosen among a set
of $n$ values, say $t_1,\dots,t_n$. A measurement at time $t_k$ gives some value $a_k=\pm1$.
Under Assumption~(A1), the outcome of a measurement has a definite value even if
the measurement is not performed, so that we can define a joint probability of
form $\rho(a_1,\dots,a_n|s_1,\dots,s_n)$,
where $s_k$ is a binary variable encoding the information on the actual execution 
of the $k$-th measurement. If the measurement is executed, then $s_k$ is set equal to 
$1$, otherwise $s_k$ is set equal to $0$. Under Assumption~(A2), it is clear
that
\be\label{joint_prob_distr}
\rho(a_1,a_2,\dots|s_1,s_2,\dots)=\rho(a_1,a_2,\dots).
\ee
which is analogous to Fine's condition for locality in Bell's scenario~\cite{fine}.
Let us denote by $C_{i,j}$ the correlation functions 
$\langle a_i a_j\rangle\equiv\sum_{a_1,a_2,\dots}a_i a_j\rho(a_1,a_2,\dots)$.
For $n=4$, Equation~(\ref{joint_prob_distr}) implies, among others, the inequality
\be\label{CHSH_ineq}
C_{1,3}+C_{2,3}+C_{2,4}-C_{1,4}\le 2
\ee
which is identical to the CHSH inequality~\cite{chsh}. The latter
refers to a scenario in which two separate parties each perform one of two
possible measurements. Depending on their choice, one party gets outcomes 
$a_1$ or $a_2$ and the other party outcomes $a_3$ or $a_4$. The inequality 
is easily proved by observing that $\rho$ is the convex hull of deterministic
distributions taking the values $0$ or $1$. Among these distributions, the left-hand side
of the inequality takes the maximum value $2$ with $a_i=a_j$ $\forall i,j\in\{1,2,3,4\}$.

Since a violation of CHSH inequalities cannot be classically reproduced 
without communication between the parties, this implies that the correlations
in the LG scenario cannot be classically reproduced without a flow of information from
$t_2$ to $t_3$ if Ineq.~(\ref{CHSH_ineq}) is violated. This is true 
even if the flow does not allow for signaling, that is, even if the
initial quantum state has maximal entropy. 

Let us formalize this implication in the framework of the ontological model~\ref{causal_model}
of the previous section. Before the measurement at time $t_1$, the system is
described by some ontic state $\lambda_0$ with probability distribution
$\rho(\lambda_0)$. The
system is measured at time $t_1$ or $t_2$ and the probability distribution of
the outgoing ontic state $\lambda_1$ is $\rho(\lambda_1|\lambda_0,s_1,s_2)$, where $s_1\ne s_2$.
Since only one of the two measurements is executed in each run, we can 
associate the outcomes $a_1$ and $a_2$ with a joint probability of the form 
$\rho(a_1,a_2|\lambda_0,\lambda_1)$ such that the marginals are the probabilities 
imposed by Model~\ref{causal_model}. Finally, the ontic state $\lambda_1$  
conditions the outcome of a second measurement at time $t_3$ or $t_4$ 
according to a probability $\rho(a_3,a_4|\lambda_1)$ whose marginals are
given by the model.
If there is no information flow from the first measurement to the second
measurement, then we have 
\be\label{no_flow}
\int d\lambda_0 \rho(\lambda_1|\lambda_0,s_1,s_2)\rho(\lambda_0)=\rho(\lambda_1).
\ee
\begin{lemma}
\label{no_flow_lemma}
Let us consider the LG scenario with $4$ times. If Eq.~(\ref{no_flow}) is
satisfied (no information flow), then the 
correlations satisfy the LG inequality~(\ref{CHSH_ineq}).
\end{lemma}
{\it Proof}. We have
\be\label{infe_lemma1}
\begin{array}{l}
\rho(a_1,a_2,a_3,a_4|s_1,s_2,s_3,s_4)=
\int d\lambda_0 d\lambda_1  \rho(\lambda_0)  \\
\rho(a_3,a_4|\lambda_1)\rho(a_1,a_2|\lambda_0,\lambda_1)
\rho(\lambda_1|\lambda_0,s_1,s_2).
\end{array}
\ee
By Bayes' theorem, we have
\be
\rho(\lambda_1|\lambda_0,s_1,s_2)\rho(\lambda_0)=
\rho(\lambda_0|\lambda_1,s_1,s_2)\rho(\lambda_1|s_1,s_2).
\ee
Equation~(\ref{no_flow}) implies that $\rho(\lambda_1|s_1,s_2)=\rho(\lambda_1)$.
Using these last two equations, Eq.~(\ref{infe_lemma1}) can be rewritten in
the form
\be
\begin{array}{l}
\rho(a_1,a_2,a_3,a_4|s_1,s_2,s_3,s_4)=
\int d\lambda_0 d\lambda_1  \rho(\lambda_1)  \\
\rho(a_3,a_4|\lambda_1)\rho(a_1,a_2|\lambda_0,\lambda_1)
\rho(\lambda_0|\lambda_1,s_1,s_2).
\end{array}
\ee
Integrating over $\lambda_0$, we have
\be
\begin{array}{l}
\rho(a_1,a_2,a_3,a_4|s_1,s_2,s_3,s_4)=
\int d\lambda_1  \rho(\lambda_1)  \\
\rho(a_3,a_4|\lambda_1)\rho(a_1,a_2|\lambda_1,s_1,s_2).
\end{array}
\ee
Since the correlation $C_{1,2}$ does not appear in the
LG inequality~(\ref{CHSH_ineq}), we can apply the replacement
$$
\begin{array}{c}
\rho(a_1,a_2|\lambda_1,s_1,s_2)\rightarrow 
\rho(a_1|\lambda_1,1,0) \rho(a_2|\lambda_1,0,1)\equiv \\ \rho'(a_1,a_2|\lambda_1)
\end{array}
$$
without changing the value of the left-hand side of the inequality. Therefore,
the left-hand side can be evaluated using the modified distribution
\be
\rho'(a_1,a_2,a_3,a_4)=
\int d\lambda_1  \rho(\lambda_1)  
\rho(a_3,a_4|\lambda_1)\rho'(a_1,a_2|\lambda_1).
\ee
Since the joint probability $\rho'(a_1,a_2,a_3,a_4)$ 
is not conditioned by the execution of the measurements,
the LG inequality is satisfied. $\square$.

\subsection{Quantum violation of the LG inequalities}
Let us show that quantum theory violates the LG inequality with four times $t_1,\dots,t_4$.
The unitary evolution is taken time-independent with Hamiltonian equal to
\be
\label{hop_hamil}
{\hat H}=|1\rangle\langle -1|+|-1\rangle\langle 1|
\ee
The evolution over a time interval $\Delta t$ is described by
the unitary operator
\be
\hat U(\Delta t)=\mathbb{1}\cos \Delta t -i {\hat H} \sin\Delta t,
\ee
so that the correlation between $a_k$ and $a_l$ at times $t_k$ and $t_l$ is
\be
\langle a_k a_l\rangle=\cos 2 (t_k-t_l).
\ee
The left-hand side of the LG inequality is maximal at $t_{k+1}=\pi/8+t_k$ with
$k\in\{1,2,3\}$. The maximum is the {\it Tsirelson bound} $2\sqrt{2}$, which 
violates the ``macrorealistic'' bound $2$.

Thus, an ontological model of these four measurements must exhibit some
information flow from the past to the future (Lemma~\ref{no_flow_lemma}).
Is this information necessary for every value of 
$t_1$ and $t_2$ on Alice's side? To answer this question,
let us find the values of $t_1$ and $t_2$ such that the inequality is violated
for some $t_3$ and $t_4$. Maximizing the left-hand side of the LG inequality
with respect to $t_3$ and $t_4$, we get the value
\be
2 \left( |\cos(t_2-t_1)|+|\sin(t_2-t_1)|\right),
\ee
which always violates the LG inequality, apart from the values $t_2=t_1+m_1 \pi/2$, $m_1$
being an integer. These values correspond to the case in which the measurements
at time $t_1$ and $t_2$ project on the same basis. Thus, whenever the two measurements
do not commute, there must be some finite amount of communication from Alice to
Bob. That is, Eq.~(\ref{no_flow}) does not hold. Thus, we have the following~\cite{montina_isit}.
\begin{lemma}
\label{lemma_viol}
Let us consider a qubit undergoing measurement $\hat A_1$ or $\hat A_2$. If
$\hat A_1$ and $\hat A_2$ are incompatible, then the 
probability distribution of the ontic state after the measurement depends
on which one has been executed.
\end{lemma}

\section{Information erasure}
\label{inf_erasure}
In Sec.~\ref{leg_garg_sec}, we have shown that the violation of the LG inequalities
can be reproduced in the ontological model only if there is a flow of information
from the past to the future.
This communication is implied by Lemma~\ref{no_flow_lemma}.
The fundamental aspect is that this communication is required even if 
the initial quantum state has maximal von Neumann entropy.
However, communication is possible only if 
the carrier of the information has initially a low entropy or its state can be erased
by a low-entropy external device. 
Suppose that the initial quantum state with maximal von Neumann entropy 
corresponds to maximal ignorance of the classical state. From the violation
of the LG inequalities, the execution of a measurement must be encoded in the classical 
state $\lambda$ of the system, which is the only carrier of information in the model. 
Since the classical variable $\lambda$ has initially maximal entropy, the measuring 
device has to exert an {\it information erasure} on the variable.

In Ref.~\cite{montina_isit}, we proved the {\it information-erasure} theorem
by assuming that the maximally mixed quantum state is compatible with
a uniform distribution on the ontological space. We also provided a more
intricate proof by employing the weaker assumption that there is a distribution 
$\rho(\lambda)\in\Omega$ with finite entropy associated with some
quantum state. Here, we adapt the first straithforward proof by directly employing
the second assumption. 
\begin{assumption}
\label{assu2}
There is a quantum state $\hat\rho$  compatible with a distribution
$\rho(\lambda)\in\Omega$ whose entropy is finite (not equal to $-\infty$).
\end{assumption}
This assumption holds, for example, in $\psi$-ontic models (such as the
Beltrametti-Bugajski model), in which a maximally mixed quantum state
is compatible with a uniform distribution of the ontic state.
Assumption~\ref{assu2} can also be justified by some results in quantum 
communication complexity. It is known that a process of quantum state 
preparation of a qubit and subsequent measurement can be simulated by a finite 
amount of classical communication~\cite{toner}. Although no generalization 
to $n$ qubits with one-way communication is known, it is likely that 
such a generalization exists. A finite classical communication implies
that there is a Model~\ref{causal_model} that is (completely) 
$\psi$-epistemic~\cite{montina_comm_epist0,montina_comm_epist}, 
the mutual information between $\lambda$ and the quantum state being
finite. A direct way to obtain a finite mutual information is to impose 
Assumption~\ref{assu2} for every quantum state. 

\begin{lemma}
\label{lemma_rho_max}
Under Assumption~\ref{assu2},
there is a unique distribution $\rho_{max}(\lambda)\in\Omega$ with finite
maximal entropy associated with $\hat\rho_{max}$. This distribution is
invariant under unitary evolutions.
\end{lemma}

The converse of the statement of this lemma 
would be that every distribution associated
with $\hat\rho_{max}$ has entropy equal to $-\infty$. For example,
consider the closure of the space of distributions whose support
is contained in a countable set of points with one accumulation
point (convex hull of Dirac delta distributions). All these 
distributions would have entropy equal to $-\infty$. Conversely,
if all the points of the space are accumulation points, the
closure contains smooth distributions with finite entropy.
The central point of
the proof is to show that there is at least one distribution with finite
entropy, which trivially follows from Assumption~\ref{assu2}.

{\it Proof of Lemma~\ref{lemma_rho_max}}. 
Let $\rho(\lambda)$ be a distribution in $\Omega$ with finite 
entropy associated with the quantum state $\hat\rho$. There is a statistical
mixture of unitary evolutions that transforms $\hat\rho$ to $\hat\rho_{max}$.
Since the transformation does not decrease the entropy, there is a distribution
$\rho_0\in\Omega$, associated with $\hat\rho_{max}$, that has also finite
entropy. Define $\bar\Omega$ as the set of distributions associated
with $\hat\rho_{max}$.
Since this set contains at least one distribution with finite entropy -- 
namely, $\rho_0(\lambda)$ -- and by Property~1 of Model~\ref{causal_model}
the entropy of distributions in $\bar\Omega$ is upper-bounded, it follows that
there exists a distribution $\rho_{max}(\lambda)\in\bar\Omega$ with
finite maximal entropy. By convexity of $\bar\Omega$ and strict concavity
of the differential entropy,
the distribution is unique and, thus, it is invariant under unitary evolutions.
$\square$.

A direct consequence of this lemma is the following.
\begin{theorem}
\label{lemma_erasure}
The entropy of the distribution $\rho_{max}(\lambda)$ associated with 
a qubit is decreased  by executing a nontrivial measurement $\hat A$.
\end{theorem}
{\it Proof.} Let us prove that a measurement $\hat A_1$ erases information
in Model~\ref{causal_model}. Let the initial probability distribution be
$\rho_{max}(\lambda)$. A second measurement $\hat A_2$ is defined by
some unitary evolution and subsequent measurement of $\hat A_1$ such
that $\hat A_1$ and $\hat A_2$ are incompatible. Let us assume that
measurement $\hat A_1$ does not erase information. Thus, the outgoing
probability distribution has maximal entropy, that is, it is 
equal to $\rho_{max}(\lambda)$. Since unitary evolutions preserve the
distribution $\rho_{max}$,
also measurement $\hat A_2$ has outgoing distribution $\rho_{max}$.
But this is in contradiction with Lemma~\ref{lemma_viol}. We conclude
that a measurement performs an information erasure on the 
distribution $\rho_{max}$ by decreasing its entropy. $\square$

The Beltrametti-Bugajski model provides
an illustration of the theorem. As discussed in Appendix~\ref{sec_dBB}, 
an infinite amount of information
is erased by a measurement in this model. However, there are less demanding
classical models of qubits in which the erased information is finite. 
In Ref.~\cite{montina}, it was shown that the erasure of just one bit suffices
to account for the outcome statistics of a two-state system, the measurements 
being performed at two arbitrary times.

This theorem does not directly imply that every finite sequence
of measurements must erase information at some stage of the process. However,
it is possible to prove the following.
\begin{theorem}
\label{gen_theorem}
If the initial entropy of the ontic state is finite, there is a finite 
sequence of measurements such that information is erased at some stage of
the process.
\end{theorem}
This theorem has been proved in Ref.~\cite{montina_isit} (Theorem~3 therein).

\section{Interpretation of information erasure}
\label{sec_interpr}

The partial reset of the ontic state upon a measurement can be interpreted
as an ontological relic of the quantum-state update following a measurement.
This reset has a direct
relation to a recent debate on $\psi$-epistemic theories~\cite{pbr,barrett}.
One of the motivations for these theories is to explain the quantum-state collapse 
after a measurement as a gain of information about the system. Whereas in $\psi$-ontic 
theories this collapse leads to the erasure of infinite information, in a 
$\psi$-epistemic theory a measurement should just lead to a gain of information
of the present state of affairs of the system, so that we should expect no information 
erasure once the measurement outcome is forgotten. At most, we could expect an increase 
of entropy given by stochastic kicks of the measuring device. The fact that a measurement 
must erase some amount of information suggests that at least part of the
quantum-state collapse is ontic and not epistemic -- the reset of the quantum state is 
mirrored by a partial reset of the classical state.  Similarly,
the results in Ref.~\cite{barrett} somehow show that the ontic 
state must hold more information about the quantum state than 
a maximally $\psi$-epistemic theory suggests. However, it is also 
interesting to note that information erasure is displayed in 
any dimension, whereas the results in Ref.~\cite{barrett} hold 
in dimension greater than $3$. Indeed, the Kochen-Specker model
of a qubit is maximally $\psi$-epistemic.

Information erasure can have a justification once we
consider the overall process behind a measurement. No measurement
is possible if some external system with lower entropy is not available.
For example, it is impossible to see what is inside a cavity 
through a small hole if the electromagnetic radiation in
the cavity is in thermal equilibrium with the internal surfaces.
A measurement device can be modeled as a pointer at some
rest position and getting entangled with the measured system 
after an interaction. For the sake of simplicity, suppose
the system and the pointer are each a qubit. The system
is initially in the superposition 
$|\psi\rangle_S=\alpha |0\rangle_S+\beta |1\rangle_S$, whereas
the pointer is in the rest position $|0\rangle_P$. We want
to model a measurement projecting into the basis $\{|0\rangle_S,|1\rangle_S\}$.
The vectors in this basis do not evolve during the
measurement. Thus, the interaction generally transforms 
the states $|0\rangle_S|0\rangle_P$ and $|1\rangle_S|0\rangle_P$ 
into $|0\rangle_S|0\rangle_P$ and
$|1\rangle_S|1\rangle_P$, respectively (CNOT gate). Thus, the overall state
becomes $\alpha |0\rangle_S|0\rangle_P+\beta |1\rangle_S|1\rangle_P$,
which eventually undergoes decoherence induced by the environment.
Thus, the outcome of the measurement is encoded into the 
state of the pointer.
This modeling of a quantum measurement does not work
if the initial state of the pointer is completely unknown. 
Thus, the device can be seen as a kind of `low temperature' bath 
that `cools' the system during the measurement
with a transfer of entropy from the latter to the former. 

There is an interesting consequence of information erasure.
In a theory of spontaneous collapse of the wave-function, the 
entropy of the system generally decreases during a collapse.
If the wave-function is taken as part of the ontology, the 
decrease is even infinity, as discussed in Appendix~\ref{sec_dBB}.
Assuming that the entropy of the overall universe cannot decrease,
we could wonder where this lost entropy ends up. Personally,
we embrace the point of view that the unitary evolution always holds
for a closed system like the whole universe. Thus, information
erasure never occurs in closed systems, but it is induced by
the environment. This suggests some 
speculations on preparation contextuality, which has a relation with information 
erasure.

\subsection{Preparation contextuality}

A quantum measurement can be realized with different experimental procedures,
but some details of the implementation are actually irrelevant for improving 
our prediction of the outcomes. These details are called the {\it context}
of the measurement. In quantum theory, the Hermitian operator associated with
the measurement summarizes the essential aspects of an experimental procedure. 
The assumption that the context keeps being irrelevant in any underlying
ontological theory leads to {\it no-go} theorems, such as Kochen-Specker
and Bell theorems.

In Ref.~\cite{spekkens}, Spekkens extended the notion of contextuality
to the case of state preparation. Like for measurements, 
some of the details of the preparation
of a quantum system do not improve our predictions on the outcomes 
of any subsequent measurement, they are the {\it context} of the
preparation.
The quantum state $\hat\rho$ summarizes all the relevant information 
on the preparation.
Employing this generalized notion of {\it context},
Spekkens showed that any ontological rephrasing of quantum theory is 
{\it preparation contextual}.
Namely, there are mixed quantum states whose associated probability
distribution $\rho(\lambda)$ on the ontological space depends
on the preparation context. For example, there are infinite ways
for representing a maximally mixed state $\hat\rho_{max}$ as
convex combination of pure states. In a non-contextual ontological
theory, these different representations should correspond to the same
distribution $\rho(\lambda)$. This turns out to be false. Indeed, 
information erasure is an example of preparation contextuality.
Suppose that a qubit is in the  maximally mixed state $\hat\rho_{max}$.
Theorem~\ref{lemma_erasure} states that there is a probability
distribution $\rho(\lambda)$ associated with $\hat\rho_{max}$ such
that a measurement $\hat A$ transforms $\rho(\lambda)$ to
a different distribution with lower entropy. 
Since we trace out the outcome, the quantum state after the measurement 
is still $\hat\rho_{max}$.
Thus, we have two preparation procedures which are operationally
identical, but generate different distributions on the ontological
space. In
one procedure, we take a maximally mixed quantum state and we
do nothing else. In the second procedure, we take the maximally mixed
state and execute the measurement $\hat A$ (the outcome being ignored). 
We get the same quantum state, but different distributions of $\lambda$. 

It is worth to stress that information erasure occurs only in open
systems under our assumptions. More generally, we can
state that preparation contextuality occurs only in open
systems.
Indeed, every scenario in which the preparation only involves
the choice of a unitary transformation of some given initial
quantum state is non-contextual. In particular, this is the case
when only pure quantum states are considered. In general, an
ontological theory is made non-contextual with respect to
`unitary preparations' by setting a suitable initial probability
distribution of the ontic state. Namely, the preparation is non-contextual if
the initial distribution is invariant with respect to 
transformations that do not change the initial quantum
state.  General preparations can always be implemented by choosing only 
unitary evolutions on the system and an ancilla. 

\section{Time symmetry}
\label{sec_time_symm}
In this section, we discuss the relation between information erasure
and the breaking of time symmetry in ontological theories.
In Ref.~\cite{leifer}, Leifer and Pusey introduced an operational
definition of time symmetry and proved that time symmetry is in
conflict with causality under a plausible assumption that they call
$\lambda$-mediation. This assumption is also employed in our
Model~\ref{causal_model} and states that all the information
about the previous manipulations on the system  is encoded into
the ontic state $\lambda$ of the system.

The essence of their argument is caught by this simple scenario.
There is an experimental procedure $\omega$ on a qubit defined by an initial 
quantum state $\hat\rho$ and two sets of measurements $\{\hat A_1,\hat A_1'\}$
and $\{\hat A_2,\hat A_2'\}$.
Alice first executes one of the two measurements $\hat A_1$ 
and $\hat A_1'$. Subsequently, Bob executes another measurement chosen 
between $\hat A_2$ and $\hat A_2'$. Let us denote by $r_A$ and $r_B$
the choices of Alice and Bob, respectively, and by $a_A$ and $a_B$ 
their respective outcomes. The procedure generates $a_A$ and $a_B$
according to some conditional probability $\rho_\omega(a_A,a_B|r_A,r_B)$.
\begin{definition}
Given two procedures $\omega_1$ and $\omega_2$, if
\be
\rho_{\omega_1}(a_A,a_B|r_A,r_B)=\rho_{\omega_2}(a_B,a_A|r_B,r_A),
\ee
then one procedure is called the time reverse of the other.
\end{definition}
The operational procedures defined by Leifer and Pusey are slightly
different, as the first party executes a state preparation, whose
protocol is quite tricky. The
fact that our model of measurement has an incoming and outgoing
ontic state enables us to consider procedures which are 
more symmetric in their execution, with two measurements instead
of one preparation and one measurement 

In general, a procedure does not have a time reverse. This is made
clear by the fact that Alice can signal to Bob, but not vice-versa.
Thus, quantum theory seems intrinsically time asymmetric. However,
as argued by Leifer and Pusey, this asymmetry should not be considered
as fundamental, but a consequence of the low-entropy state at the
beginning of the universe. In the respect of our two-party procedure,
a low-entropy initial quantum state 
enables Alice to manipulate the qubit and signal to Bob.
To get a more symmetric procedure, we have
to prepare the qubit in the maximally mixed state $\hat\rho_{max}$,
so that a measurement made by Alice cannot transfer information
on the qubit. This is analogous to a one-time pad cryptographic protocol,
the cryptographic key being the initial pure quantum state, which 
is actually unknown. With this choice, every procedure turns out to have
a time reverse. The ratio of this state preparation is to wash out the asymmetry
due to the lower entropy of the initial state of the universe. It
is worth to stress that this strategy does not actually make the procedures
completely time symmetric. Indeed, a measurement in its very essence always
requires initial low entropy of some part of the measuring device,
as pointed out in Sec.~\ref{sec_interpr}. Without some low-entropy
object, it is not possible to define a measurement. The fact that
Alice and Bob can perform measurements requires some degree of
time asymmetry, which however is not directly attached to the
system. This point will be fundamental in the discussion of the
result of Leifer and Pusey. It will also provide a hint for eluding
information erasure and the break of time symmetry.

Let us define the main assumption used in Ref.~\cite{leifer}.
\begin{assumption} (Time Symmetry).
\label{time_symm}
If a procedure $\omega_1$ has a time reverse $\omega_2$, then
there is a  Model~\ref{causal_model} and a bijective map 
$\lambda\rightarrow f(\lambda)$ such that the map transforms a
process of $\omega_1$ into a time-reverse process of $\omega_2$
within the model.
\end{assumption}
The map $f(\lambda)$ is not generally the identity. Consider for example
the time-reverse transformation in classical mechanics for which 
the direction of the momentum is inverted. Leifer and Pusey do not
impose conditions on the map $f$ other than bijection. It is reasonable
to assume also volume-preservation, that is, entropy-preservation, but
this is not necessary in their proof.
Time symmetry is a kind of `no fine tuning principle'; if a quantum process
is time-symmetric at the operational level, the principle would require 
that the symmetry is inherited at the ontological level.

In the defined framework, let us prove that {\it time symmetry} leads
to a contradiction. Our proof is slightly different from Leifer
and Pusey proof, but it is essentially equivalent.
\begin{lemma}
\label{lemma_LP}
Every Model~\ref{causal_model} leads to a contradiction under
the assumption of time symmetry.
\end{lemma}
{\it Proof}. Let us prove it by contradiction. Given a procedure $\omega_1$
on a qubit,
let $\hat\rho_{max}$ be the initial quantum state. Alice performs one
of two incompatible measurements. From Lemma~\ref{lemma_viol}
the outgoing probability distribution depends on the executed measurement.
By time symmetry, there is a simulation in which the distribution
before Bob's measurement depends on his choice. But this breaks
causality. Thus, time symmetry leads to a contradiction. $\square$
\newline
Assuming time symmetry as a fundamental property of physics, Leifer
and Pusey conclude that the contradiction is removed by dropping
causality. 

In other words, causality implies a break of the time symmetry
and, thus, of the `no fine tuning principle'. Let us show that
information erasure is the mechanism leading to this break. As 
pointed out previously, the procedure with maximally mixed initial 
quantum state employed in Lemma~\ref{lemma_LP} is not completely
time symmetric, because Alice and Bob need low-entropy ancillary
states to execute a measurement. If we watch to the details of
the overall procedure, we identify $4$ different execution
times. Alice sets the pointer of the measurement device to the 
rest position at some time $t_0$. She executes
a measurement by letting the device interact with the system
at time $t_1>t_0$. At time $t_2>t_1$, Bob sets
the pointer of his device to the rest position.  Finally,
he executes a measurement at $t_3>t_2$. 

If we revert time,
we get a completely different process, in which Alice and
Bob set their pointers after the execution of their measurements.
Including the pointers in the description of the overall 
process, the procedure does not have a causal reverse
procedure. Thus, there is no `a priori' reason for assuming
that the system satisfies the time-symmetry assumption.

Although the initial state of the system has
maximal entropy, this is not true for the measuring device.
Through the device, Alice can erase information by decreasing
the entropy and, thus, encode information on the executed
measurement into the state $\lambda$ of the system. This enables 
her to influence the outcomes
of Bob's measurements. This influence goes from the past to
the future, so that Bob cannot influence Alice's outcomes.
Information erasure is a fine-tuned mechanism, since it does not
allow a party to signal toward the future. This mechanism leads
to the break of the time symmetry.

It is worth to note that Bell's theorem already implies a break
of causality if Lorentz invariance is assumed at the ontological
level. Thus, the tension between the `no fine-tuned principle' and
causality is already displayed by nonlocality under the assumption
of Lorentz invariance.

\section{Splitting into parallel coexisting realities}
\label{many-worlds_sec}
In Ref.~\cite{leifer}, it has been  argued that the many-worlds (MW) 
theory fails to satisfy the time-symmetry Assumption~\ref{time_symm}. 
Here, we actually show that the relaxation of the hypothesis of 
single actual outcomes allows for evading the Leifer-Pusey theorem.

Using the hybrid approach of Ref.~\cite{montina_branch},
which combines the branching of MW theory with the randomness 
of single-world ontological theories, we propose a symmetric model 
that simulates the outcomes of two consecutive measurements on a qubit. 
Our model circumvents the information-erasure theorem and the theorem 
of Leifer and Pusey by transferring the information flow and time asymmetry 
to the measuring devices and the subsequent comparison of results,
which inherently involve time-asymmetric processes. This model is
the temporal counterpart of the local model presented in 
Ref.~\cite{montina_branch} simulating spatial correlations.

\begin{figure}[h]
\centering
\includegraphics[width=3.5in]{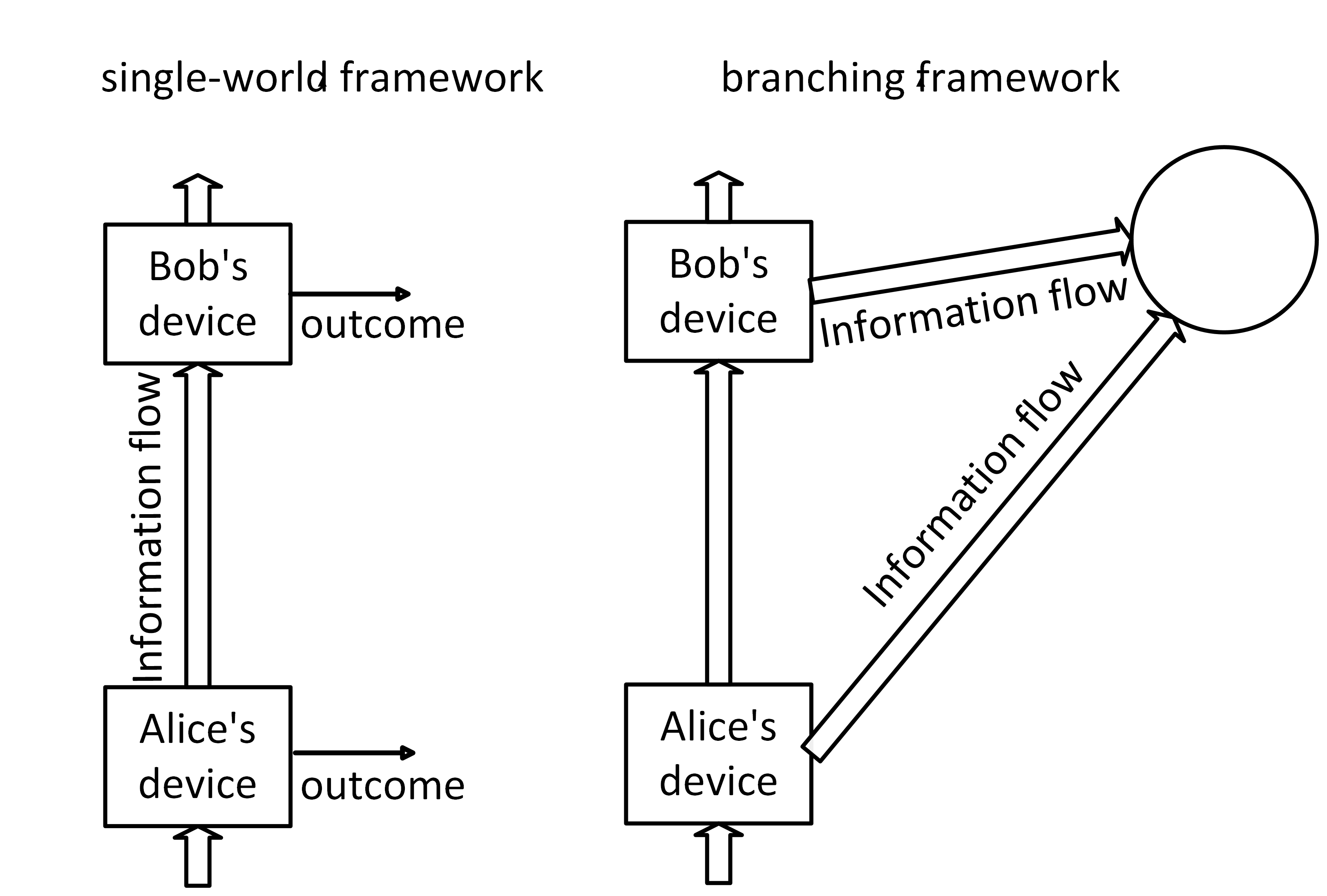}
\caption{Information flow in both single-world and branching frameworks. In the
first case, information erasure is necessary. In the second case, the
devices and the observers are the carriers of information and no
erasure is required}.
\label{fig1}
\end{figure}

The argument for dropping the hypothesis that measurements have single, 
definite outcomes is as follows. In the scenario described in the previous 
section, we observed that the measuring device is the only physical object 
with non-maximal entropy. Therefore, the measured system can carry 
information only if some of its entropy is transferred to the device. This 
transfer leads to information erasure.
To prevent this reset of the system, one could argue that the device itself
-- and, consequently, Alice -- serve as carriers of information. Since Alice 
and Bob must eventually meet to compare their results, the information about 
Alice’s measurement may ultimately reach Bob. But how can this information
alter the outcome that Bob previously observed? Rather than assuming a sudden 
change in Bob’s memory or retro-causality, we can argue that two 
parallel realities ({\it instances}) evolve separately after each measurement. 
Eventually, these realities are properly paired at the meeting point to 
correctly reproduce the observed correlations. The information flow in both 
single-world and branching frameworks is illustrated in Fig.~\ref{fig1}.

The model illustrating this idea is as follows.
The two parties are now allowed to perform any projective
measurement on the qubit. Alice and Bob's measurements are
denoted by the Bloch vectors $\vec a$ and $\vec b$, respectively.
The qubit is in a maximally mixed quantum state.
At the ontological level, the qubit is described by two
unit vectors $\vec x_0$ and $\vec x_1$. The state of Alice's 
device is represented by two bits, say $s_A$ and $n_A$, which
take values $\pm1$. Similarly, $s_B$ and $n_B$ represent 
the state of Bob's device. Before the measurement, the devices 
are set in some {\it rest} state. The bits $s_A$ and $s_B$ are
set equal to the outcomes after the measurements.
As we argued previously, the measuring devices inherently break 
time-symmetry. 

Assuming parallel coexisting realities, we say that,
when Alice's device performs measurement $\vec a$,
it branches into two different alternatives, say $A_1$ and
$A_{-1}$. We assume that Alice observes outcomes 
\be\label{s_A_eq}
s_A=\text{sign}(\vec a\cdot\vec x_0)
\ee
and $-s_A$ in branches~$A_1$ and $A_{-1}$, respectively.
Then, she sets
\be\label{n_A_eq}
n_A=\text{sign}(\vec a\cdot\vec x_0)\text{sign}(\vec a\cdot\vec x_1)
\ee
in both the branches. The vectors $\vec x_0$ and $\vec x_1$ are left
untouched by the measurement, that is, no information carried by
the ontic state of the qubit is erased. This split propagates
in the space through physical systems entering in contact 
each other. In particular, the device splits, then Alice observes
the outcome and she splits too, and so on.
Bob receives the qubit and performs measurement $\vec b$.
Also Bob branches into two different alternatives $B_1$ and 
$B_{-1}$, observing outcomes 
\be\label{s_B_eq}
s_B=\text{sign}(\vec b\cdot\vec x_+)
\ee
in branch $B_1$ and $-s_B$ in branch $B_{-1}$, 
where $\vec x_+\equiv \vec x_0+\vec x_1$.
Furthermore, he sets
\be\label{n_B_eq}
n_B=\text{sign}(\vec a\cdot\vec x_+)\text{sign}(\vec a\cdot\vec x_-),
\ee
where $\vec x_-\equiv \vec x_0-\vec x_1$.
Since the results of each party have to be compared for 
estimating the correlations over many runs of the experiment, they
have to meet each other. Whenever this meeting occurs, 
the branches of each party are finally paired according to the rule
\be
\begin{array}{c}
(n_A,n_B)\ne(-1,-1) \Rightarrow  A_{\pm 1}\leftrightarrow B_{\pm1} \\
(n_A,n_B)=(-1,-1) \Rightarrow  A_{\pm 1}\leftrightarrow B_{\mp 1}.
\end{array}
\ee
In Ref.~\cite{montina_branch}, it has been shown that this protocol,
which is a variant of Toner-Bacon model~\cite{toner},
reproduces exactly the quantum predictions by taking one of the two
merged branches at random. This model needs no information erasure,
since the devices, which are necessarily in an initial low-entropy 
state, are the carriers of information.

The branching {\it \`a la} many-worlds theory has be exploited 
in Ref.~\cite{montina_branch} to provide a local model of quantum
correlations with finite information flow. In the present scenario,
it could be argued that the branching is not necessary, since
Alice and Bob have a time-like separation. In a single-world
scenario, we could imagine that Alice's device sends the bit $n_A$ 
to Bob's device and conditions the outcome $s_B$. However, what
would be the carrier of this information if the two parties
never interact through some physical medium
before Bob executes his measurement?
In a branching framework,
on the side of each party, the device first splits, then the
party observes the outcome and he/she also splits. Finally,
the two parties compare the results by meeting together and
the branches of each one are suitably paired. The information
is always carried by some physical system, namely the devices
and the observers.
It is worth to note that the two devices employ different rules
in the generation of their outcomes. However, this asymmetry is
not attached to the measured system. Anyway, it 
can be easily
removed by adding a random bit that establishes if the parties
must use the decribed protocol or its time-reversal version.

Since both information erasure and the breakdown of time symmetry are 
forms of preparation contextuality, this model suggests that multiple 
parallel events, in the style of many-worlds theory, may offer a way 
to circumvent contextuality -- or at least some of its manifestations.

\section{The clumsiness loophole in Leggett-Garg tests}
\label{sec_clumsy}
In light of our findings, it is worthwhile to conclude by addressing 
an issue that 
arises in the experimental testing of macrorealism \`a la Leggett-Garg. 
We will not engage here in the debate about the necessity of both the 
hypotheses (A1 and A2) for defining macrorealism. A thorough discussion 
on this topic can be found in Ref.~\cite{maroney}. Let us just mention 
again that the de Broglie-Bohm theory is macrorealistic but does not 
satisfy the second hypothesis. In fact, we have previously discarded 
this hypothesis by assuming that unitary evolution always holds.
In this section, however, we adopt the definition of macrorealism that 
incorporates both hypotheses. According to this definition, macrorealism 
entails a breakdown of unitarity and prohibits the superposition of 
macroscopically distinct states. Under these conditions, macrorealism 
can be experimentally disproven by demonstrating that a Leggett-Garg 
inequality is violated, as predicted by quantum theory.
Leggett-Garg inequalities are temporal analogs of Bell inequalities, 
with the hypothesis of noninvasiveness replacing Bell’s hypothesis of 
locality. As highlighted in Ref.~\cite{wilde}, this distinction makes 
Leggett-Garg~(LG) tests more susceptible to loopholes than Bell tests. 
While the locality postulate forbids any non-local influence between 
spatially separated systems, hypothesis (A2) merely asserts that 
noninvasive measurements on macrostates are possible. However, it 
does not rule out the possibility that experimental imperfections in 
the measurement process might influence the subsequent state of the system.
If an experimental test reveals a violation of the Leggett-Garg inequalities, 
this could indicate that the measurement technique introduced some noise, 
thereby affecting the system’s subsequent state and violating hypothesis~(A2). 
Ref.~\cite{schmid} distinguishes between noninvasiveness and realized noninvasiveness, 
emphasizing that experimental tests must ensure the latter. This vulnerability 
in LG tests is commonly referred to as the clumsiness loophole~\cite{wilde}.
Our findings help mitigate this loophole.

The model underlying the LG argument 
is quantum theory with a superselection rule~\cite{maroney}, where the macrostates correspond 
to quantum states. This model belongs to the class of ontological models in which 
the quantum state is elevated to the status of an ontic object. Additionally, 
one may assume the presence of supplementary ontological variables.
For a system with two macroscopically distinct states, if both 
hypotheses hold, the system should exhibit a stochastic process transitioning between 
the two states such that the LG inequalities are satisfied, unless ’clumsy’ 
measurements introduce uncontrollable noise. This raises the question: what kind 
of noise would be necessary to observe a violation of the inequalities?
Our result demonstrates that an invasive measurement alone is insufficient; 
it must also erase some amount of information, under the assumption that the ontic 
state initially has maximal entropy. Thus, the 
clumsiness loophole is relevant only in highly specific cases involving implemented 
measurements that erase information. However, a macrorealistic model in which 
non-ideal measurements lead to information erasure appears highly improbable. In 
general, uncontrollable noise tends to increase entropy rather than decrease it.
This can be stated in a different way. We previously noted that the quantum-state
collapse is a kind of information erasure. If an experiment detects a violation
of the Leggett-Garg inequalities, this implies that part of the erasure has
been performed by a measurement and not by a spontaneous collapse, as demanded
by macrorealism \`a la Leggett-Garg.
We conclude that the theorem of `information erasure'
diminishes the effectiveness of the clumsiness loophole argument. Therefore, in 
our opinion, an experimental violation of the Leggett-Garg inequalities strongly
supports a breakdown of macrorealism as defined by hypotheses (A1-A2). Furthermore,
if we assume causality and no erasure of information, then the
violation of LG inqualities would support the suggestion of the previous
section that measurements have multiple actual outcomes {\it \`a la} many-worlds
theory.

In this context, it is worth noting that a much simpler test of macrorealism arises 
from the fundamental observation that a single quantum measurement disturbs the system, 
thereby conflicting with Assumption~(A2). Specifically, measuring the position of a 
particle will scatter it, disrupting the system. For instance, observing which path a 
particle takes in a double-slit experiment destroys the interference pattern. According 
to macrorealism, if the alternative paths are macroscopically distinct, the measurement 
should not be invasive (no signaling in time). This observation leads to a test of 
macrorealism involving only two measurements at different times~\cite{brukner,halli,kofler}.
However, this scenario is not suitable for demonstrating information erasure and is thus 
more vulnerable to the clumsiness loophole. Indeed, in the `no signaling in time' approach, 
the system is prepared in a specific initial state, meaning that the initial ontic state cannot 
be assumed to be fully random, which is necessary for the proof of information 
erasure. As a result, the measurement could alter the ontic state without erasing
information. Thus, the ‘no signaling in time’ test does not prove that measurements erase 
information.

\section{Conclusions}

Assuming that a quantum process of multiple projective measurements
is described by an underlying stochastic process over an ontological
space of states of realities (ontic states), we have previously 
shown that the interaction of a system with a measuring device erases 
the information carried by the system~\cite{montina_isit}. This suggests 
that the quantum-state update after a measurement cannot be entirely epistemic.
Information erasure can be interpreted as a flow of entropy from the system 
to the measuring device and, indeed, it is not displayed if the device
is included in the description. Here, we have provided a simple
proof by assuming that there is quantum state that is compatible with
a probability distribution of finite entropy. 

We have then discussed the proof of Leifer and Pusey~\cite{leifer} that 
causality implies a break of time symmetry. We have identified information 
erasure as the mechanism breaking this symmetry. Indeed,
a measuring device is inherently time asymmetric, since
it needs to be set into some initial rest state. During 
the measurement, its low entropy is used to erase information,
so that the time asymmetry is trasferred to the measured system.
The strangeness of this process is that it is a fine-tuned mechanism, 
since it does not allow for signaling through the measured
system.

Since communication requires a carrier with non-maximal entropy -- and 
initially, only measuring devices serve as suitable carriers -- 
we are tempted to infer that the information flow from the
past to the future passes through the device (and the
observer), rather than the system. This would avoid the entropy 
transfer from the system to the device and, thus, information erasure.
However, providing a causal description of consecutive measurements 
with devices and observers as carriers forces us to drop 
the assumption that measurements have single, definite outcomes.
A similar conclusion was drawn by Deutsch and Hayden from the
assumption of locality~\cite{deutsch2}. Motivated by this insight, we have
introduced a model, inspired by the many-worlds theory, 
that simulates the outcomes of two consecutive arbitrary projective 
measurements on a qubit in a maximally mixed quantum state. This model 
does not require information erasure  and is thus completely time-symmetric. 
It is a temporal version of a model recently introduced in 
Ref.~\cite{montina_branch},
which simulates quantum correlations without nonlocal influence.
Notably, although the set of allowed measurements is infinite, our model 
requires a finite information flow. Moreover, it uses only 
two ‘unweighted’ coexisting realities 
(called {\it instances} in Ref.~\cite{montina_branch}), 
meaning that they have equal probability
of being experienced.
This contrasts with the many-worlds theory, 
where branches are weighted by amplitudes, leading to interpretative 
issues. The model can reproduce the quantum probabilities of 
the outcomes with unweighted instances because of its stochasticity.
Note that the probability distribution of the pair of outcomes 
$1/4(1-s_A s_B \vec a\cdot\vec b)$ is not uniform. To reproduce it
with a unweighted counting in a deterministic model would require
infinite instances. Randomness allows to reproduce the quantum
probabilities even with one instance (like in all single-world
ontological models such as de Broglie-Bohm theory). However,
one instance is not enough for having a time-symmetric model
as implied by Leifer-Pusey theorem. We have shown that two
instances suffice to circumvent the theorem in the case of two
consecutive measurements on a one qubit, even if the full set of 
projective measurements is considered.
See Ref.~\cite{montina_branch} for a more detailed discussion on
the general framework of theories with multiple coexisting 
realities and the role of randomness in this framework.
Since information erasure is a kind of preparation contextuality,
we have argued that the drop of the hypothesis of single actual outcomes
could also elude contextuality or, at least, some of its manifestations.

We have concluded with a discussion on how the problem of the clumsiness loophole 
in an experimental Leggett-Garg test of macrorealism is mitigated by the 
information-erasure theorem. An experimental detection of the violation
of Legget-Garg inequalities cannot be explained through a device perturbation
adding noise to the measured system. If we reject the assumption that
devices can erase information and assume causality, the violation would support 
the hypothesis that the outcomes have multiple actual values. We have also 
argued that `no-signaling in time' tests~\cite{brukner,halli,kofler} of 
macro-realism are more vulnerable to the clumsiness loophole than the original 
Leggett-Garg test.

In perspective, it may be useful to quantify the minimal amount of
information that a measurement must erase in an ontological model or
the amount of information flow in a model {\`a la} many-worlds, as
well as the minimal number of parallel realities in time-symmetrical 
models of more general quantum systems~\cite{montina_branch}.
These further studies can have relevance in quantum communication
complexity.

\section{Acknowledgments}
This work was supported by Swiss National Science Foundation (SNF) under
the project number 182452 and Hasler Foundation under project number 23023.

\appendix
\section{Beltrametti-Bugajski model and de Broglie-Bohm theory}
\label{sec_dBB}

We discuss two examples, the Beltrametti-Bugajski model~\cite{beltrametti}
and the de Broglie-Bohm theory, to illustrate the theorem on information
erasure.
In the former, measurements erase infinite information.
The case of the de Broglie-Bohm theory is more intricate and the effect of
the measurement depends on what is considered as the measured system. If
the universal wave-function and the particle positions of the system are
taken as the underlying classical state of the system,
then a measurement does not erase information.
This happens because the system can carry an arbitrarily large amount of
information and, thus, does not satisfy our assumptions. 

The Beltrametti-Bugajski model provides the simplest illustration of 
Theorem~\ref{lemma_erasure}. In this model, the pure quantum 
state represents the ontology and, thus, is identified with $\lambda$. 
We can say that the model is just quantum theory with the wave-function
interpreted as real. Taking the Haar 
measure induced by unitary transformation, the distribution 
$\rho_{max}(\lambda)$ of maximal ignorance 
is a constant function with finite entropy. After performing a measurement
with eigenstates $\psi_1,\dots,\psi_n$, the distribution becomes
\be
\rho(\lambda)=\sum_k n^{-1} \delta(\lambda-\psi_k),
\ee
which has negative infinite entropy. Thus, an infinite amount of information
is erased by a measurement in this model. As previously remarked, there are
`cheap' classical models of qubits in which the erased information 
is finite~\cite{montina}.

The case of the de Broglie-Bohm theory is more intricate and it is necessary
to define what is considered as classical state of the measured system.
The theory is not separable, that is, a quantum system does not have its
own classical state, but it shares a global wave-function with the whole
universe. Thus, the ontic state $\lambda$ should contain this global
state. The positions of the particles are  the only private part of the ontology.
As an alternative, we can do as follows. Before a 
measurement, the measured system and the measuring device have a separable
state, say $|\psi\rangle_S |0\rangle_D$, so that the system has its
own private state $|\psi\rangle_S$. After the measurement, the quantum
state has evolved to an entangled state, say 
$|\psi_0\rangle_S |0\rangle_D+ |\psi_1\rangle_S |1\rangle_D$.
If the measurement is irreversible and no subsequent measurement will
detect the superposition of the two pointer states, then we can just
forget the measuring device and attach one of the two quantum states
$|\psi_0\rangle$ and $|\psi_1\rangle$ to the system with suitable
probability weights. Proceeding in this way, we get again the feature
displayed by the Beltrametti-Bugajski model, that is, a measurement
can erase an infinite amount of information.

Now, let us consider the case in which the global wave-function is
part of the ontology of the measured system. Considering a 
Stern-Gerlach experiment on a $1/2$-spin particle, the ontology
of the system also includes the position and momentum 
of the particle. 
The de Broglie-Bohm theory shares a nice feature with classical
mechanics. Whereas there is no Hamiltonian associated with the 
overall dynamics
of the wave-function and the particles, this dynamics preserves
the volume in the ontological space. The dynamics of a particle
is described by a time-dependent Hamiltonian with a quantum potential
determined by the wave-function. Thus, the volume in the phase space
is preserved. Furthermore, also the volume in the whole ontological
space is preserved  because of the unitary evolution of the wave-function.
This implies that the classical entropy of the $1/2$-spin particle
is conserved by a measurement. This is not in contradiction with
Theorem~\ref{lemma_erasure} because Assumption~\ref{assu2}
does not hold, since the entropy of the ontic state is always 
$-\infty$ (because of the pointers). We might introduce some
small randomness in the device state so that the entropy is
finite and the measurements are very close to projective measurements
of the spin. However, for fixed accuracy, the difference between 
the minimal and maximal entropy would tend to infinity as the number of 
measurements goes to infinity.

\end{document}